\input{aipcheck}

\documentclass[
    ,final            % use final for the camera ready runs
  ]
  {aipproc}

\layoutstyle{8x11single}

\usepackage{lineno}
\begin{document}

%\linenumbers

\title{Electron/Photon identification in ATLAS and CMS}


\author{C. Charlot, for the ATLAS and CMS collaborations}{
  address={Laboratoire Leprince-Ringuet, \'Ecole Polytechnique and IN2P3-CNRS, 
                    Palaiseau, France}
%  altaddress={On behalf of ATLAS and CMS collaborations}
}

%\author{<author2>}{
%  address={<common address for author2 and author3>}
%}

%\author{<author3>}{
%  address={<common address for author2 and author3>}
%  ,altaddress={<author1 address>} % additional visiting address
%}

\begin{abstract}
 Recent studies in ATLAS and CMS experiments for the reconstruction and identification of electrons and photons
 using full Monte Carlo and test beam data are reported.
 \end{abstract}

\maketitle

%%%%%%%%%%%%%%%%%%%%%%%%%%%%%%%%%%%%%%%%%%%%
%% MAINMATTER
%%%%%%%%%%%%%%%%%%%%%%%%%%%%%%%%%%%%%%%%%%%%

\section{Introduction}
ATLAS and CMS are the two multipurpose experiments under construction
for the Large Hadron Collider (LHC) at CERN.
As the LHC is expected to produce its first collisions in the end of 2007,
both collaborations are actively preparing the data taking and in particular developing their reconstruction tools.
The detection of electrons and photons are of particular importance at the LHC as these particles intervene in the
flagship $H \rightarrow \gamma\gamma$ and $H \rightarrow $ZZ$^{(*)}\rightarrow 4e$ channels for the search
of the Standard Model (SM) Higgs. Leptons are also important in many SUSY scenarii
as produced in the leptonic decays of charginos and neutralinos. Electrons and photons also appears in 
searches for  TeV resonances that may come from new symmetries
or as consequences of scenarii involving extra spatial dimensions. Last but not least, electrons appear in the final state of many standard model processes involving top quarks or electroweak bosons, that  constitute backgrounds to new signals or are intended to be used as calibration processes.

\section{The ATLAS and CMS detectors}
\begin{figure}
  \includegraphics[viewport=-40 0 980 655,width=0.495\textwidth,height=.3\textheight]{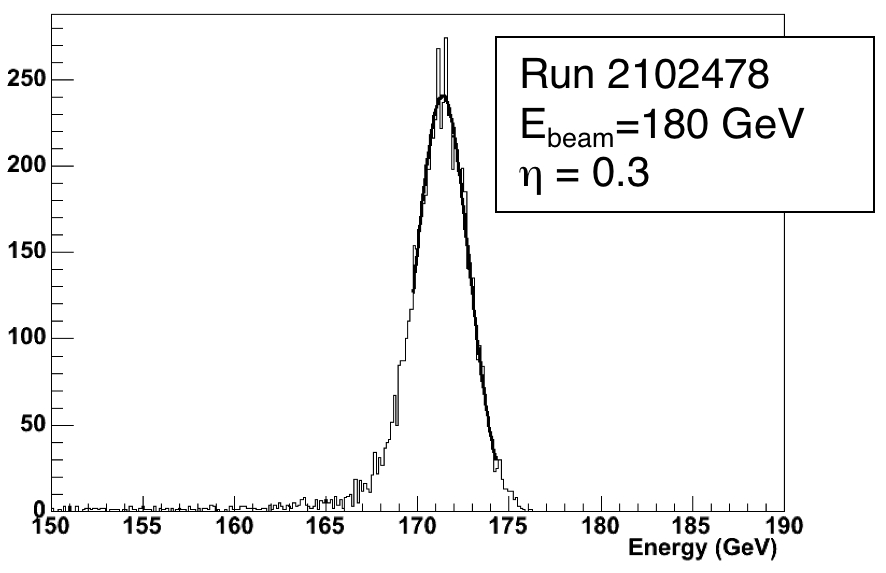}
  \includegraphics[viewport=-20 0 600 600,width=0.495\textwidth,height=.3\textheight]{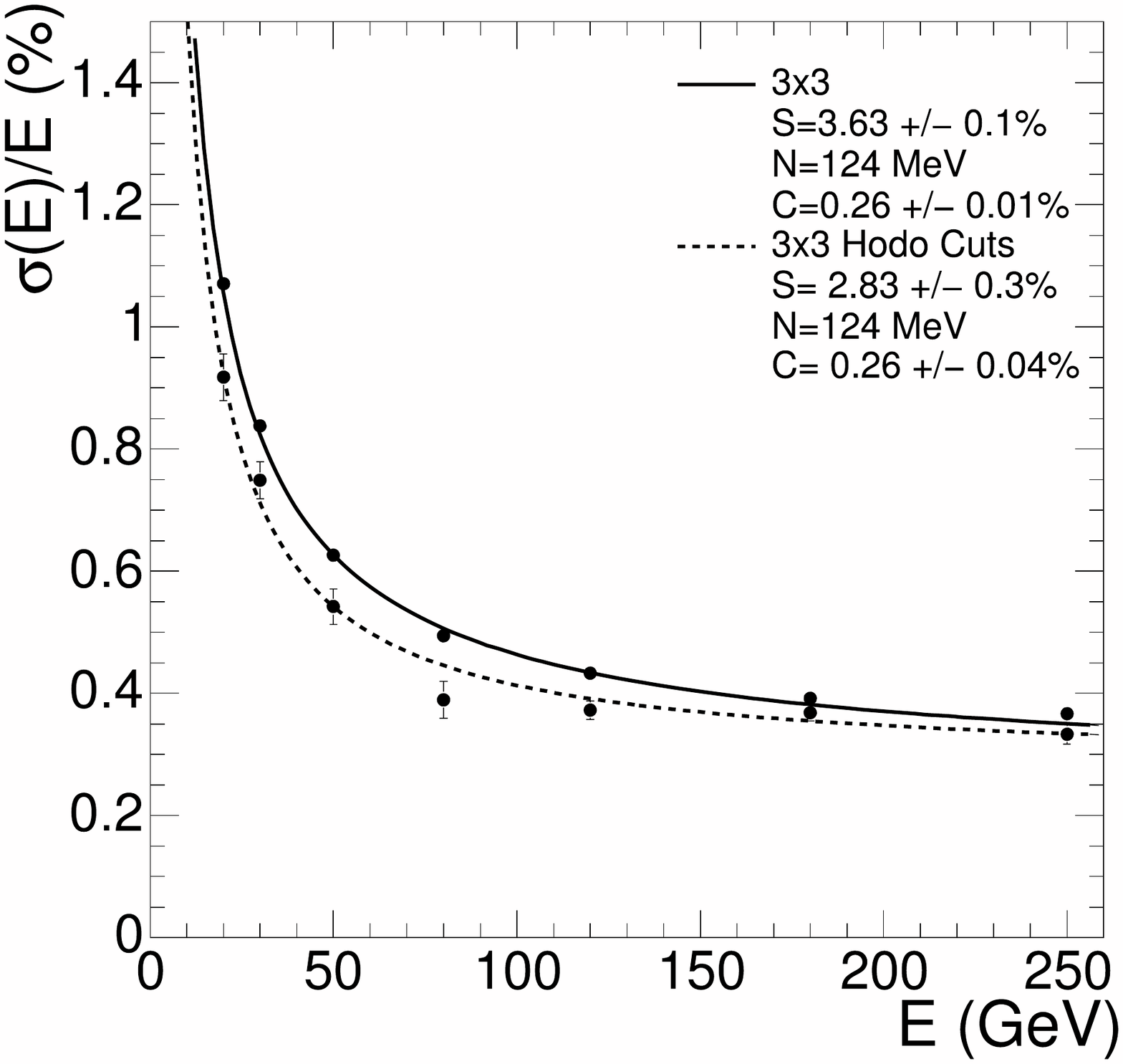}
  \label{fig:resol}
  \caption{Energy resolution of the  ATLAS and CMS electromagnetic calorimeters as measured from test beams. (left) Energy distribution with the LAr calorimeter for a 180 GeV incident energy. (right) Energy resolution as a function of the incident energy with the PbWO4 calorimeter (figure from \cite{CMSTDR}).}
\end{figure}

The ATLAS detector is built around a set of large superconducting air-core toroid magnets providing a strong bending power for the measurement of muons. The toroids are complemented by a thin solenoid of 2T surrounding the inner cavity in which the tracking system is located. The detection of electrons and photons relies in particular on the LAr electromagnetic sampling calorimeter which is located after the solenoid and before the muon spectrometer and which provides angular coverage up to $| \eta|$=2.5 for precision measurement. Of particular importance for the particle identification, the electromagnetic calorimeter is highly segmented with a 3-fold granularity in depth and an $\eta \times \phi$ granularity of %$0.003 \times 0.1$ in the front, $0.025 \times 0.025$ in the middle, and $0.05 \times 0.025$ in the back compartment, with totaling a thickness of 24X$_0$ in the barrel part  and of 26X$_0$ in the endcap parts. A presampler with a fine granularity in $\eta$ is located before
$0.003 \times 0.1$, $0.025 \times 0.025$, and $0.05 \times 0.025$  respectively in the front,  middle and back compartment, totaling a thickness of 24X$_0$ in the barrel part  and of 26X$_0$ in the endcap parts. A presampler with a fine granularity in $\eta$ is located before
the cryostat and the coil, enabling to correct for the corresponding dead material effects. The inner tracker consists of a combination of discrete semi-conductor pixel and strip detectors in the innermost part and of continuous straw tubes in the outermost part providing further particle identification capability through the production of transition radiation. More details on the ATLAS detector and its performances can be found in \cite{ATLASTDR}.
%The performances of the electromagnetic calorimeter are also reported in \cite{ATLASTDR}. 
Test beam measurements show a resolution of the LAr electromagnetic calorimeter of $(10.0 \pm 0.1) \% / \surd E \oplus (0.21 \pm 0.03)\%$ well matching the requirements. 
%Recent test beam results shown on Fig.~\ref{fig:resol}(left) demonstrate a precision of 0.83\% for 180 GeV/c incident electrons. 
Recent test beam results for high energy showers shown on Fig.~\ref{fig:resol}(left) confirm this resolution with a precision of 0.83\% for 180 GeV/c incident electrons.
The uniformity of the response within a module is measured to be within $0.37\%$ (rms). 

The CMS detector is based on a compact magnet delivering a 4T solenoidal field. The electromagnetic calorimeter, located inside the solenoid, consists of an homogeneous calorimeter made of 75848 PbWO4 scintillating crystals. The coverage for precision measurement extends up to $|\eta|=2.6$. The crystal size correspond to an $\eta \times \phi$ granularity of approximately $0.0174 \times 0.0174$ in the barrel part of the calorimeter which extends up to $|\eta|=1.479$. A preshower detector is located just before the endcap parts. The crystals are mounted in a quasi-projective geometry with a small angle (3$^{\circ}$) between the crystal axis and the line to the nominal vertex in both $\eta$ and $\phi$ directions. The crystal length corresponds to 25.8X$_0$ in the barrel and 24.7X$_0$ in the endcap parts of the calorimeter. The CMS inner tracker is made of 10 layers of silicon strip detectors 
%and 3 additional pixel layers in the innermost part (for the barrel case, and 2$\times$2 endcaps disks for the forward regions). 
complemented in the innermost part by a set of pixel layers (3 additional layers in the barrel part and 2$\times$2 endcaps disks in the forward regions).
The inner tracker provides a coverage of up to $|\eta|$=2.5. More details on the design and performances of the CMS detector can be found in \cite{CMSTDR}. Recent tesbeam results show an excellent energy resolution of the PbWO4 calorimeter
as a function of the incident energy, as illustrated in Fig.~\ref{fig:resol}(right).  For central incidence, an energy resolution of $2.8\% / \surd E \oplus 0.125 / E \oplus 0.30\%$ is measured \cite{CMSRESOL}. 

\section{Electron and photon reconstruction}

%\subsection{Electron and photon energy measurement}
The reconstruction of electron and photon objects starts by the detection of clusters in the electromagnetic calorimeters. 
Both collaborations are considering the use of fixed windows for the definition of photon clusters and of fixed windows or topological clusters in the case of electrons. For photons, a 5$\times$5 matrix is typically used (the number of cells in the ATLAS case refering to the middle compartment) while for electrons asymetrical windows or topological clustering algorithms are used. Indeed, in the case of electrons, the energy emitted by the bremsstrahlung radiation in the tracker material has to be recollected to minimize the effect of lateral fluctuations.

\begin{figure}
  \includegraphics[viewport=5 0 570 405,width=0.495\textwidth,height=.3\textheight]{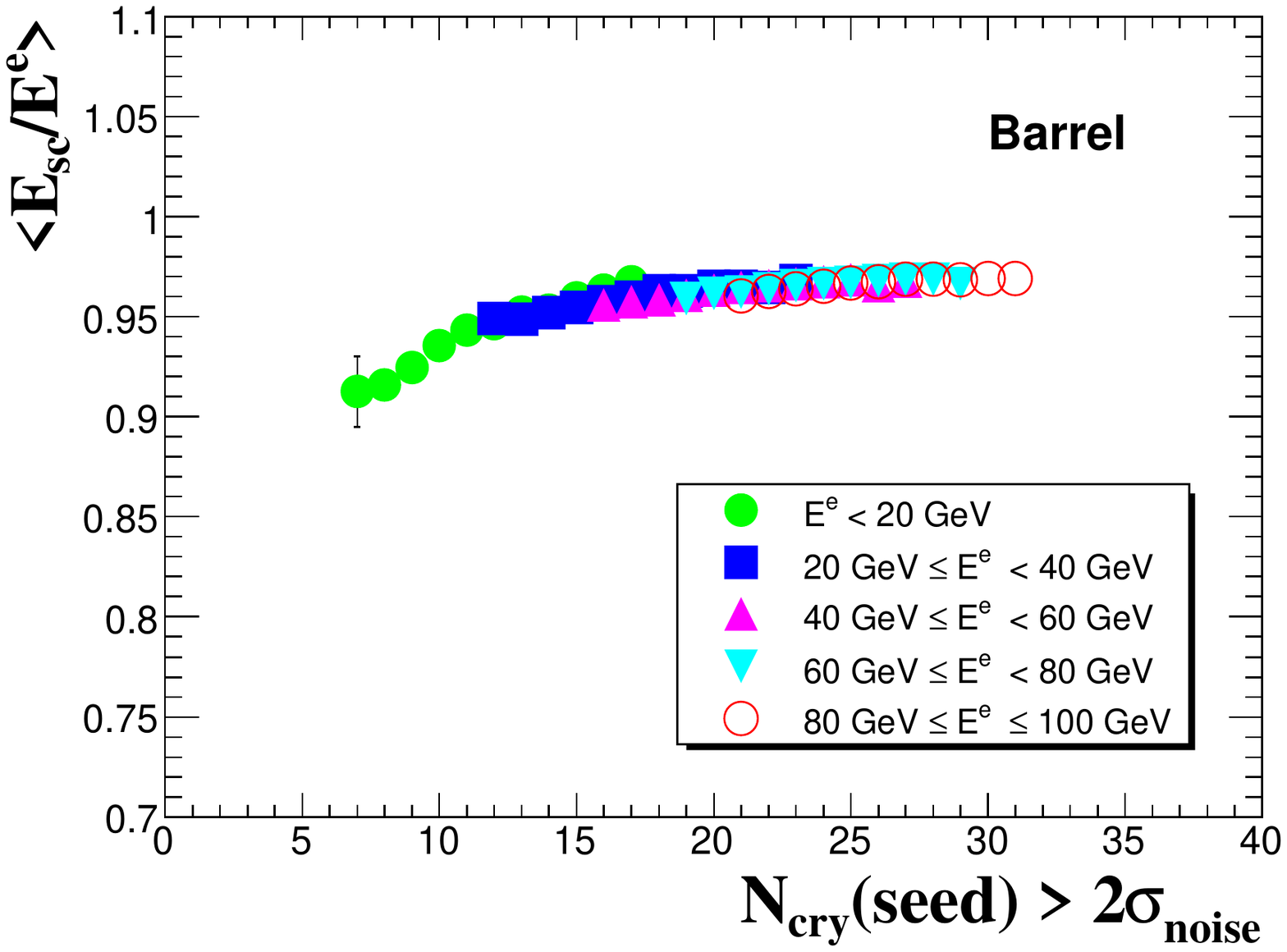}
  \includegraphics[viewport=5 0 610 445,width=0.495\textwidth,height=.3\textheight]{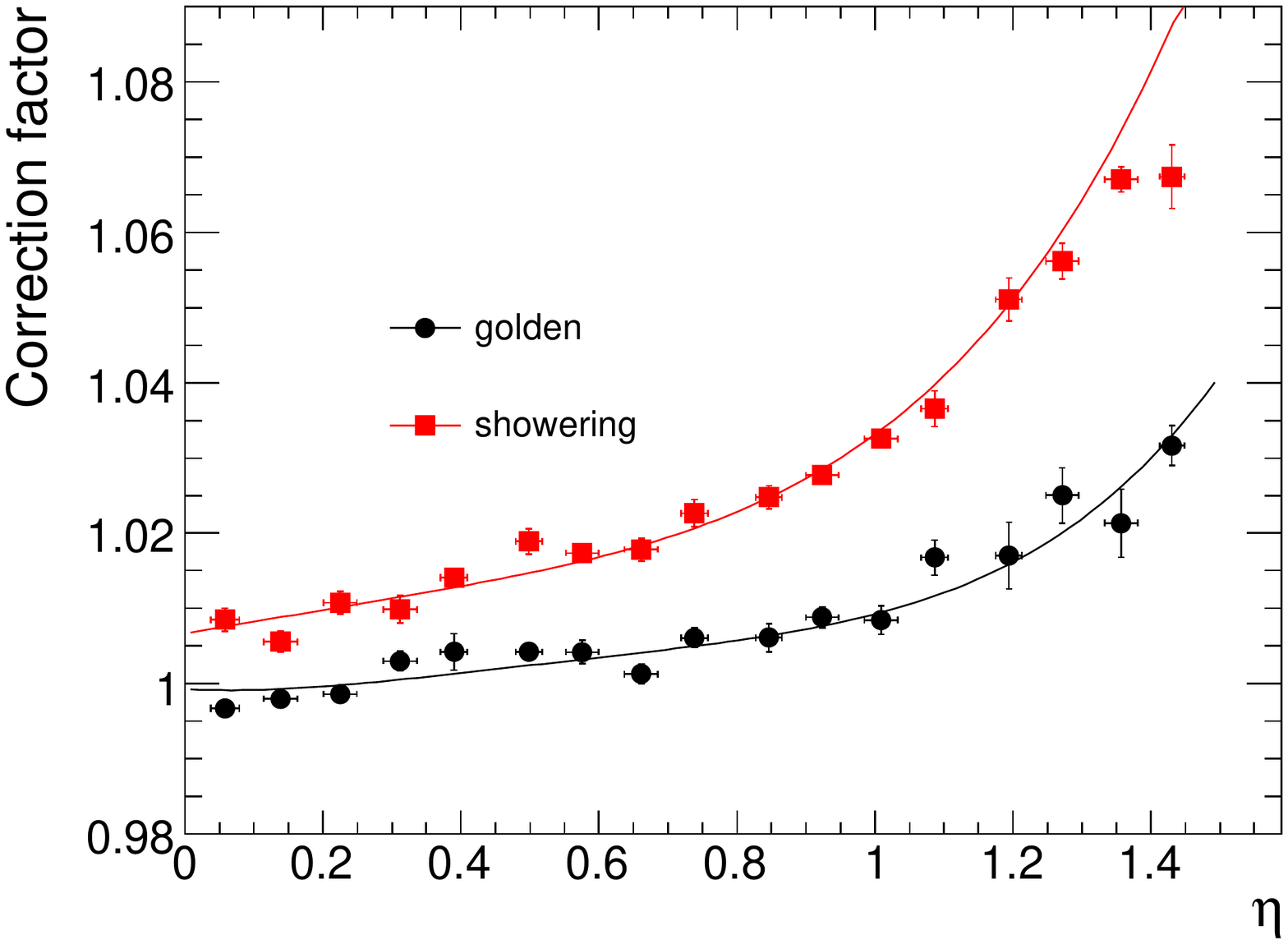}
  \label{fig:escale}
  \caption{Energy corrections. (left) Measured energy in CMS ECAL barrel as a function of the number of clusterized crystals above the noise (figure taken from \cite{CMSNOTEELE}). (right) Comparison of the $\eta$ dependent corrections functions used in CMS and derived from single electron Monte Carlo simulation (lines), and the ones (symbols) obtained using $Z \rightarrow ee$ data (figure taken from \cite{CMSNOTEZEE}).}
\end{figure}

Detailed investigations of the calorimeters response in the test beam are used to characterize the constructed calorimeter modules and study the many effects that need to be corrected for in order to reach the required precision. A full deconvolution of the various effects is carried out by the ATLAS experiment using the test beam and Monte Carlo simulations. The clusterized energy of (intercalibrated) cells is corrected for longitudinal and lateral leakage, dead material upstream of the calorimeter (making use in particular of the presampler  measurement $E^{Vis}_{PS}$), bremsstrahlung losses, and the variations of the response due to impact point,  according to:. 
\begin{equation}
E^{rec} = [a(E)+b(E) \times E^{Vis}_{PS} + c(E)(E^{Vis}_{PS} \times E^{Vis}_{1} )^{0.5} + d(E) \times 
\sum E^{calo}_{i}] \times (1+f_{leak}(depth)) \times f_{brem}(E) \times f_{cell impact}
\end{equation}
where the energy dependent correction factors are extracted from the simulation and checked with the test beam. The variation and losses of containment in CMS are corrected for by a single function of the number of clusterized crystals times an $\eta$ dependent function parametrizing the energy lost in dead material and residual $\eta $ dependent effects. Figure~\ref{fig:escale}(left) presents the containment variations for electrons in the CMS barrel, as obtained from simulations. It is expected that the corrections will be finally tuned using $Z \rightarrow ee$ real data and that the Monte Carlo will be only used to extrapolate these corrections toward kinematical regions not accessible as toward  lower $p_T$ (down to $\sim$10 GeV/c). This is illustrated on Fig.~\ref{fig:escale}(right) which presents a comparison between the $\eta$ dependent correction factors obtained with the simulation of single electrons and using $Z \rightarrow ee$ simulated data in CMS.

\begin{figure}
  \includegraphics[viewport=5 0 570 415,width=0.495\textwidth,height=.3\textheight]{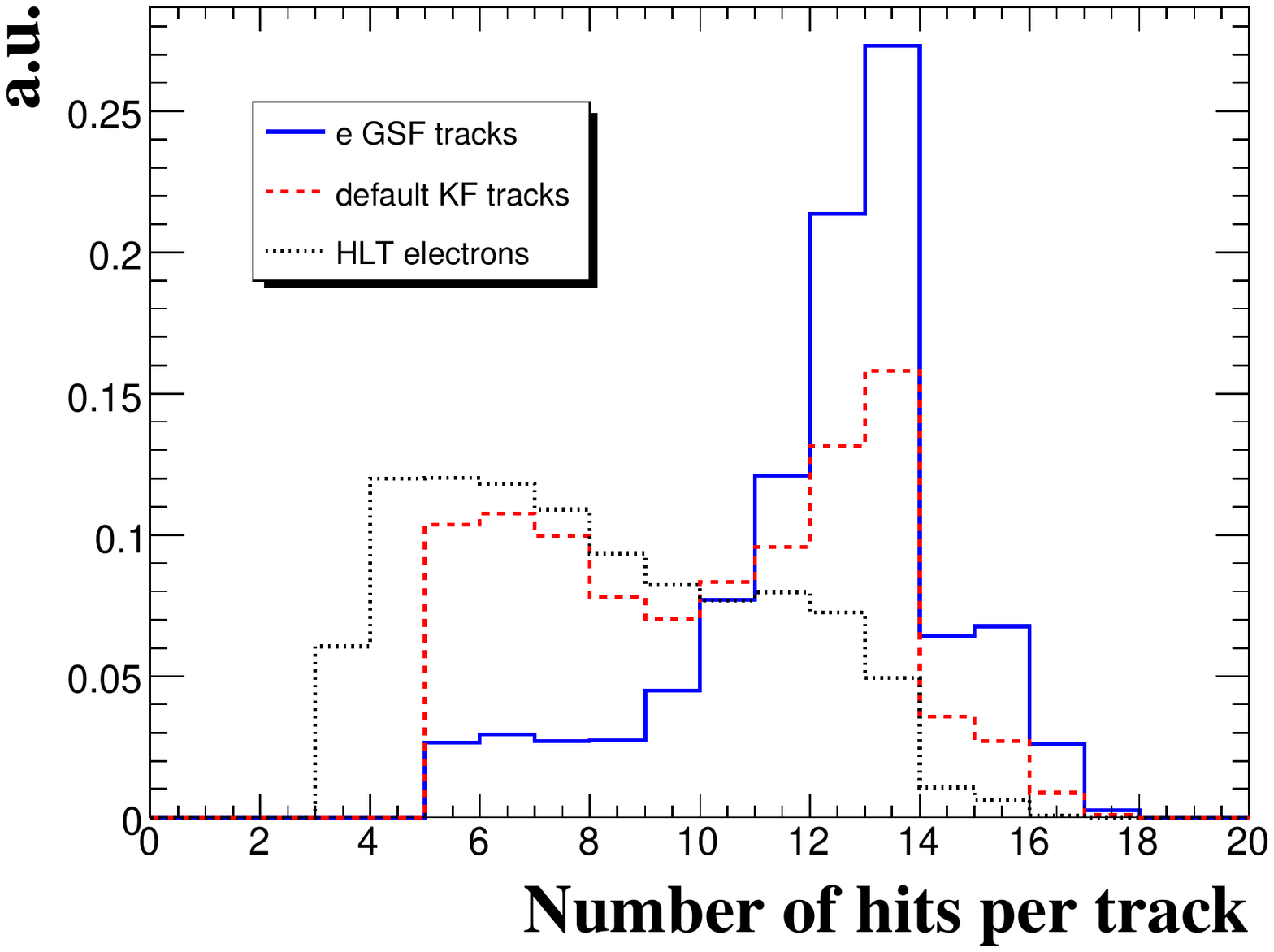}
  \includegraphics[viewport=5 0 570 415,width=0.495\textwidth,height=.3\textheight]{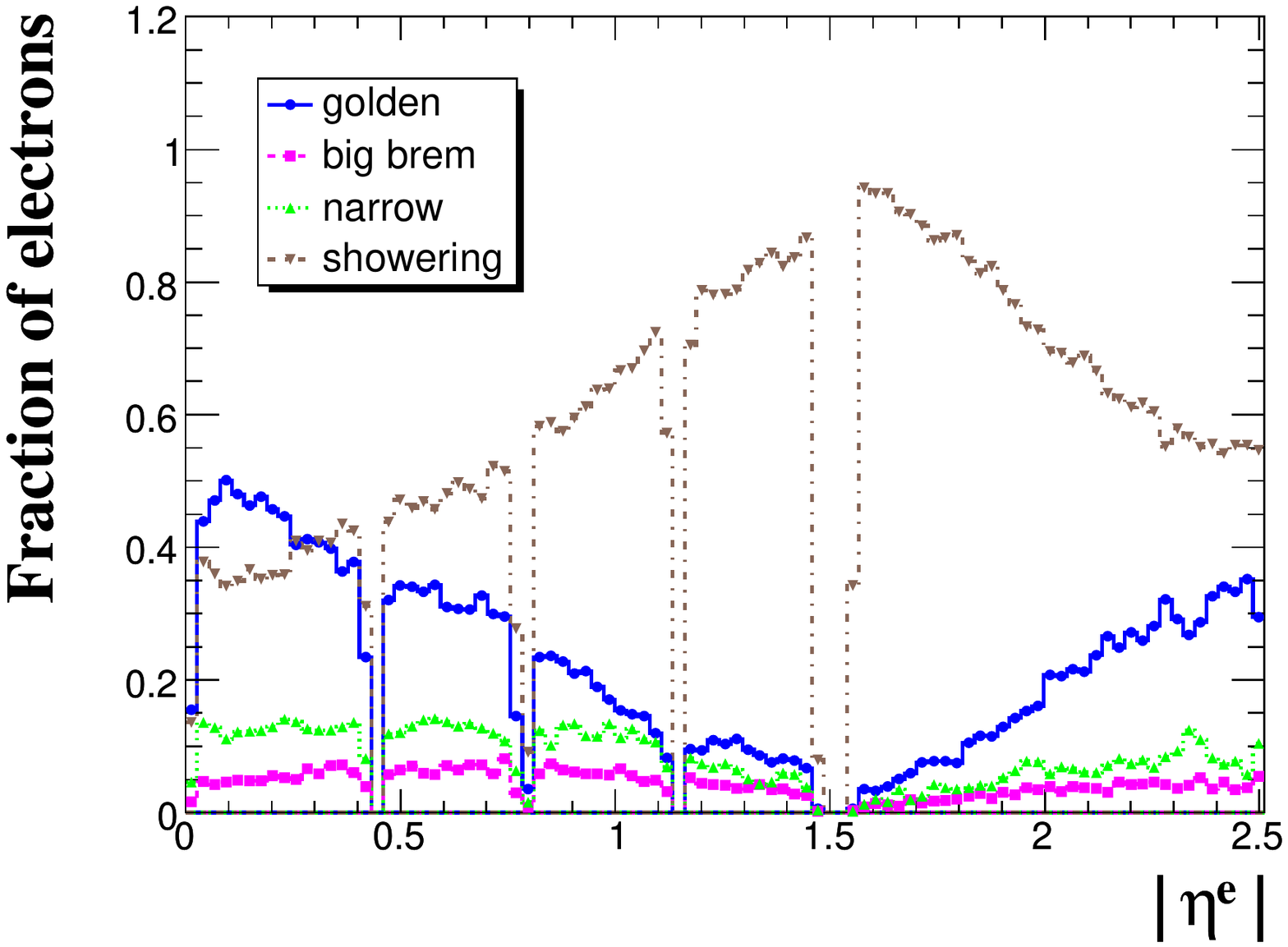}
  \label{fig:gsftracking}
  \caption{GSF tracking in CMS. (left) Number of collected hits. (right) Fraction of electron population in the different classes. The classification is based in particular on the bremsstrahlung fraction (see text).
 Figures taken from \cite{CMSNOTEELE}. }
\end{figure}

%\subsection{Electron tracking}

Electrons are then characterized by the presence of a charge track pointing to the electromagnetic cluster, on the contrary  to photons which are characterized by the absence of a matching track. Starting from the reconstructed clusters, roads can be built to search for hits in the innermost part of the tracker. This procedure is used in CMS to define electron objects at the trigger level as well as for the offline reconstruction, where looser settings and thresholds are used to efficiently reconstruct electrons down to $p_T \sim$ 10 GeV/c. As the bremsstrahlung radiation severely affects the track propagation, a dedicated tuning of the trajectory building and a Gaussian Sum Filter track fit are used in CMS. As a result, track hits can be collected up to the ECAL front face as shown on Fig.~\ref{fig:gsftracking}(left), while at the same time keeping a good momentum resolution at the vertex \cite{CMSNOTEELE}.

A further property of the GSF track fit is to provide a meaningful estimate of the track momentum at the last point. The relative difference between the momentum at the vertex and the momentum at the last point (bremsstrahlung fraction) can then be used as a measurement of the bremsstrahlung energy radiated by the electron track. Based on this measurement and on the number of observed sub-clusters, different track-cluster patterns (classes) are defined in CMS and used to derive specific corrections as well as estimates of the electron quality. The fraction of electron population in the different classes is shown on Fig.~\ref{fig:gsftracking}(right) as a function of the pseudorapidity. The fraction of badly measured ({\it showering}) electrons follows the material budget of the inner tracker, while the fraction of well measured ({\it golden}) electrons is anti-correlated with the material budget distribution.
Furthermore, as the bremsstrahlung fraction is related to the amount of material before the calorimeter, it can be used to control this amount from real data, in addition to usual variables sensitive to the integral amount of material as E/p. A recent simulation study in CMS shows that a 2$\%$ precision could be obtained on X/X$_0$ \cite{CMSNOTEELE}.

The photon reconstruction is being studied in details using the test beam, for which incident electrons
in the absence of magnetic field and dead material are close to unconverted photons. Specific corrections, in particular for the containments variations with respect to the incident position are derived from the test beam \cite{CMSGAMMA}.
Photon reconstruction is also severely hampered by the presence of dead material before the calorimeter. The conversions can be reconstructed using an inward track finding with an efficiency of $\sim$80$\%$ up to transverse radii of $\sim$65 cm in CMS. The reconstructed conversion tracks can then be fitted and used to determined the photon momentum as well as the primary interaction vertex. A precision of $\sim$1.7 mm on the longitudinal vertex coordinate is obtained for conversions produced at a transverse radius $r_T$ within 15 < r$_T$ < 58 cm in the central part of the CMS detector \cite{CMSNOTEGCONV}. 

\section{Electron and photon identification}

Further electron and photon identification is achieved using isolation, specific detectors such as the transition radiation tracker (TRT) and the highly segmented first ECAL compartment in ATLAS or the preshower detector in front of the endcap part of the CMS electromagnetic calorimeter, and finally using shower shape variables and, in the case of electrons, refined track-cluster matching.

\begin{figure}
  \includegraphics[viewport=0 10 430 625,width=0.495\textwidth,height=.3\textheight]{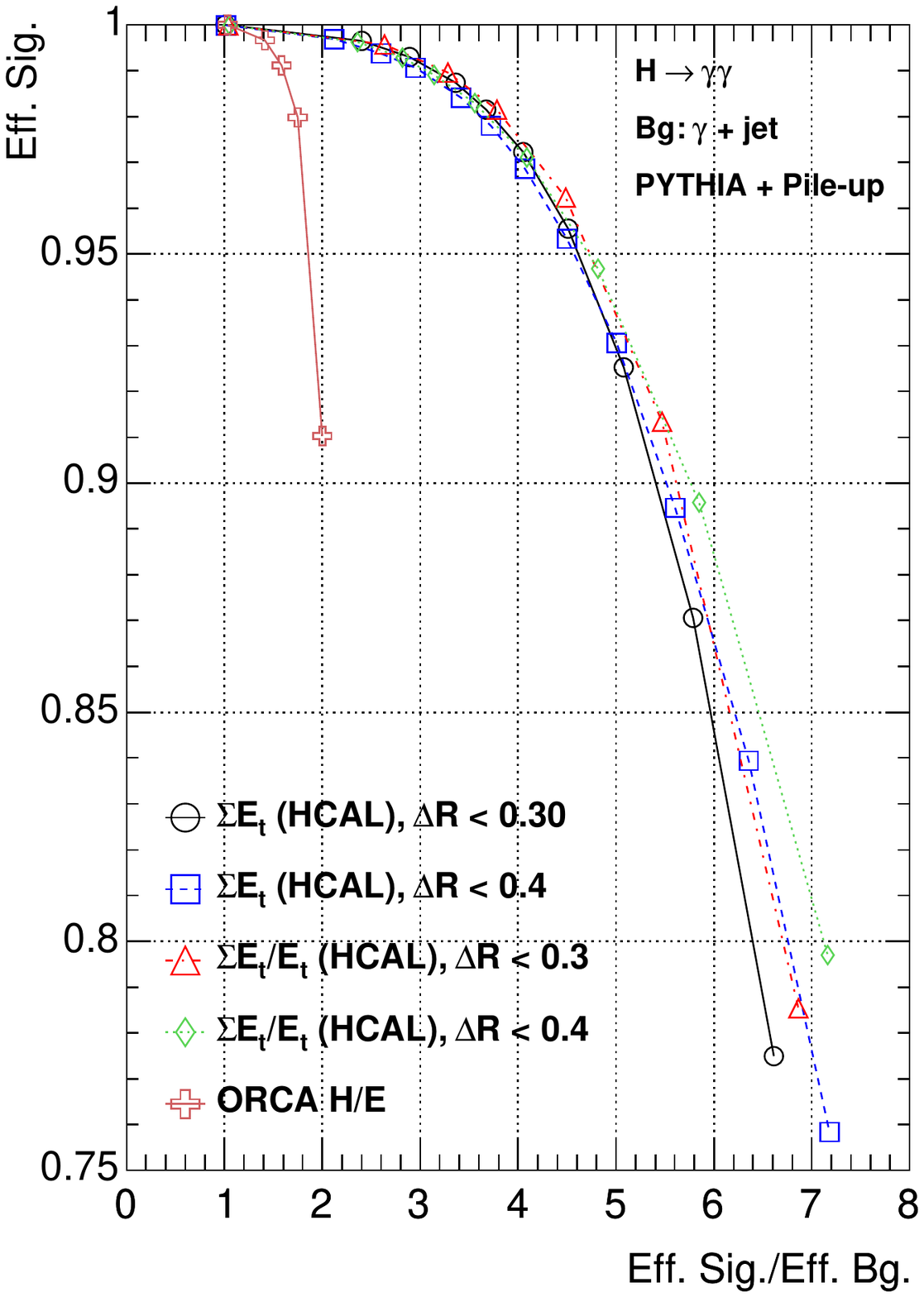}
 \includegraphics[viewport=10 27 570 395,width=0.495\textwidth,height=.3\textheight]{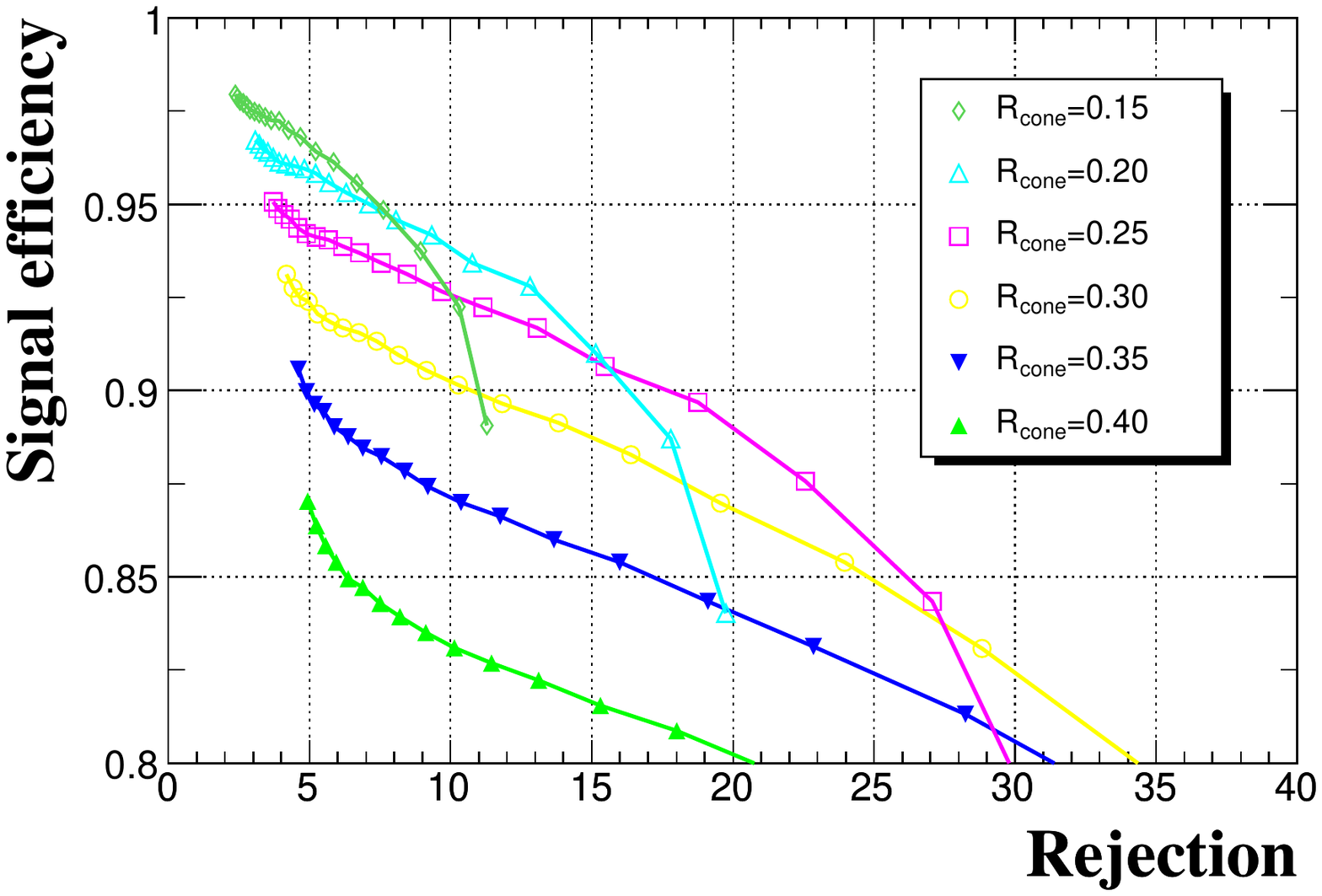}
  \label{fig:isol}
  \caption{Isolation performances. Efficiency rejection curves for (left) rejection of the $\gamma$+jet background in 
  $H \rightarrow \gamma\gamma$ using HCAL based isolation and (right)  rejection of the ${\rm t} \bar{{\rm t}}$ background in the $H \rightarrow 4e$ decay channel using track based isolation. Figures taken respectively from \cite{CMSNOTEGG} and \cite{CMSNOTE4E}.}
\end{figure}

Isolation is a very powerful tool to reject jet backgrounds in both electron and photon identification. Figure~\ref{fig:isol} shows the isolation performances obtained in CMS for the specific cases of the rejection of the $\gamma$+jet background in the search for the Higgs boson in $H \rightarrow \gamma\gamma$ \cite{CMSNOTEGG}, and for the rejection of the ${\rm t} \bar{{\rm t}}$  background in the $H \rightarrow 4e$ decay channel \cite{CMSNOTE4E}.
Different working points can be chosen according to the needed background rejection. The results shown illustrate the performances obtained using HCAL isolation and track isolation. The combination of track based and calorimeter based isolation is also investigated \cite{CMSNOTEGG}.

\begin{figure}
  \includegraphics[viewport=-30 0 710 620,width=0.475\textwidth,height=.3\textheight]{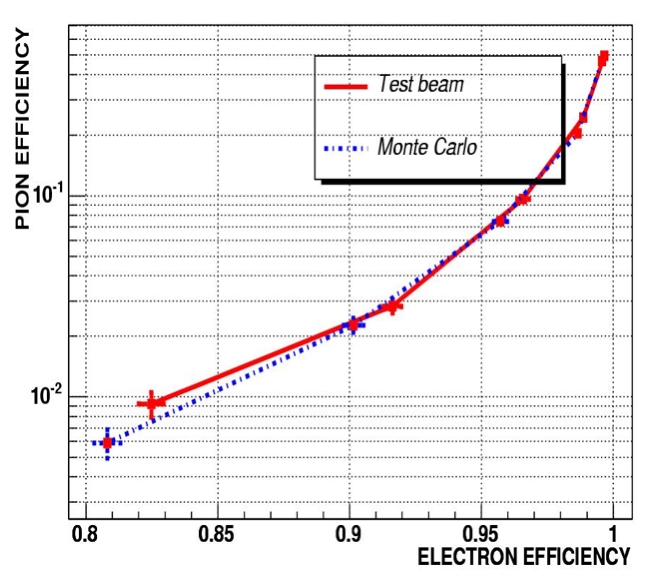}
   \includegraphics[viewport=-40 0 910 890,width=0.475\textwidth,height=.3\textheight]{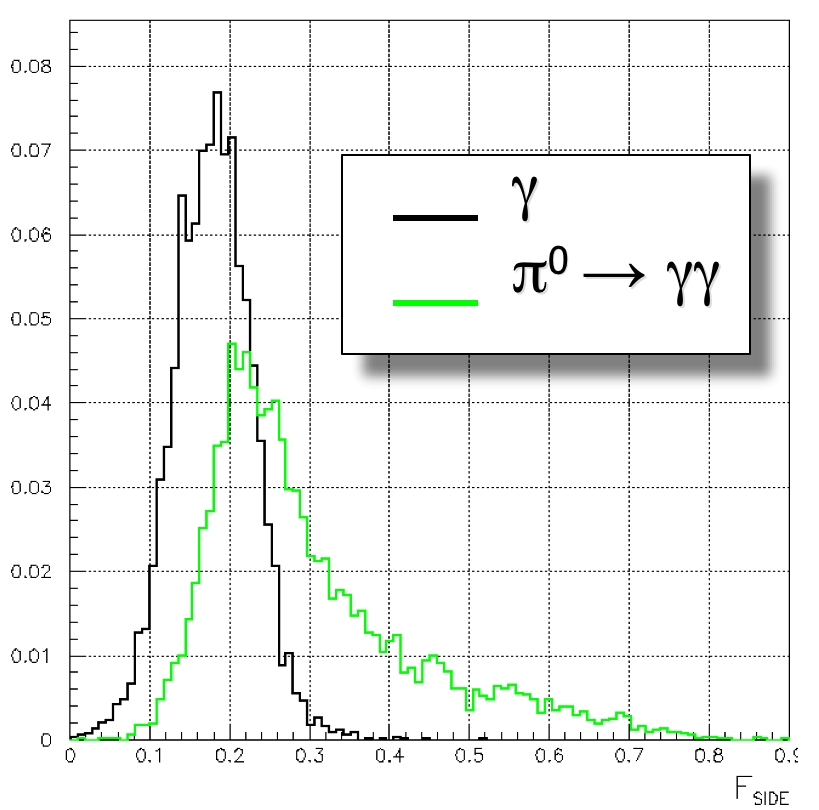}
  \label{fig:trt}
  \caption{e/$\pi$ and $\pi^0/\gamma$ separation in ATLAS. (left) pion rejection as a function of the electron efficiency for 9 GeV electrons using the transition radiation detector. (right) fraction of energy outside the shower core in the first longitudinal segment of the LAr calorimeter for $\gamma$ and $\pi^0$.}
\end{figure}

In ATLAS, the transition radiation emitted by charged particles in the outermost tracker  can be further used to separate electrons from isolated charged pions. Figure~\ref{fig:trt}(left) presents the results obtained in test beam for 9 GeV incident electrons and compared to a Monte Carlo simulation. A good agreement is observed and a rejection factor of $\sim$50 against pions is obtained for a 90$\%$ electron efficiency.

A detailed study of the shower shape in the electromagnetic calorimeter has been performed in ATLAS exploiting the high longitudinal and transversal segmentation of the detector. An excellent agreement is found between the Monte Carlo simulation and test beam data for both the mean longitudinal and transversal shapes. The transversal
shower shape, in particular in the first compartment, can then be exploited to separate photons from $\pi^0$, once isolation has been applied.  The distributions of the fraction of energy outside the shower core in the first longitudinal segment of the ATLAS electromagnetic calorimeter as obtained from Monte Carlo simulation is presented in Fig.~\ref{fig:trt}(right). A rejection factor of $\sim$3.2 against $\pi^0$ is found for a 90$\%$ photon efficiency, with good a agreement between test beam data and simulated data.

%\begin{figure}
 % \includegraphics[width=0.495\textwidth,height=.3\textheight]{figures/shapes1.eps}
 % \includegraphics[width=0.495\textwidth,height=.3\textheight]{figures/shapes2.eps}
 % \label{fig:showershape}
 % \caption{e/$\pi$ separation using the TRT in ATLAS.}
%\end{figure}

\begin{figure}
  \includegraphics[viewport=-40 0 680 660,width=0.495\textwidth,height=.3\textheight]{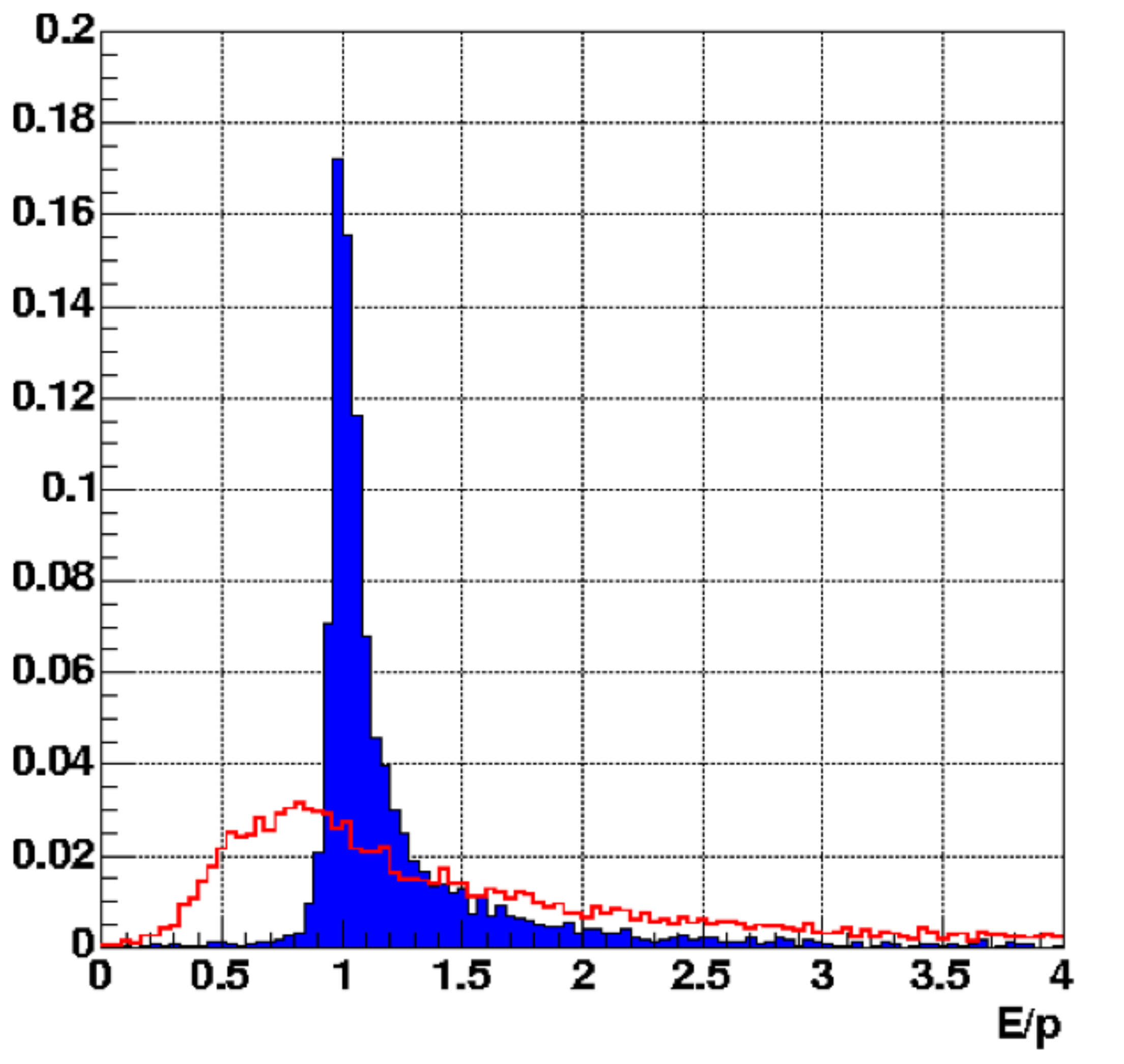}
  \includegraphics[viewport=20 10 590 410,width=0.495\textwidth,height=.3\textheight]{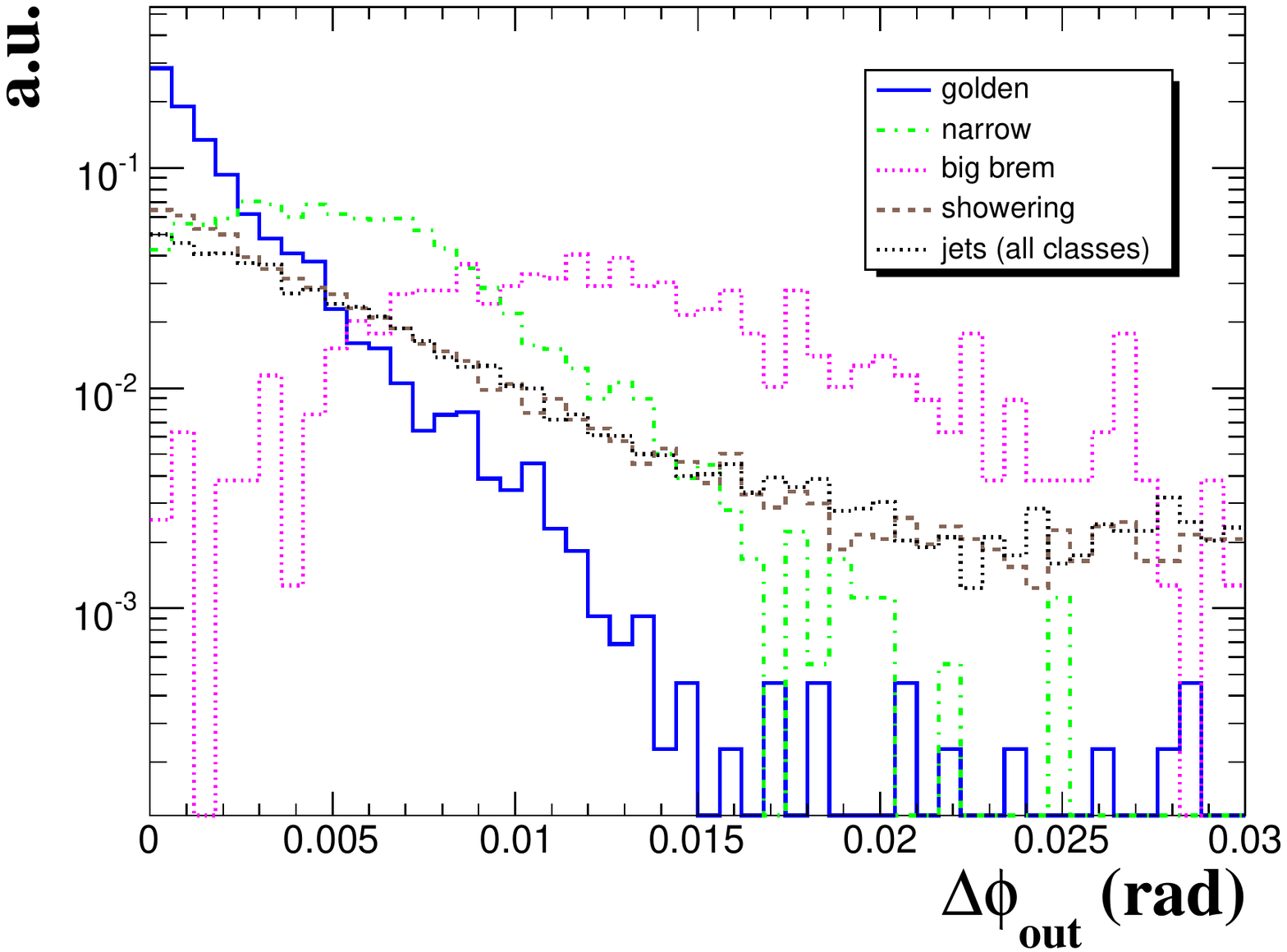}
  \label{fig:idvariables}
  \caption{Distributions of identification variables. (left) E/p distribution for electrons and jets from QCD di0jet events as expected in ATLAS. (right) Difference in $\phi$ between the cluster position and the track extrapolation from the outermost track hit for jets and for the different electron classes defined in CMS (figure taken from \cite{CMSNOTE4E}).}
\end{figure}

Refined track-cluster matching, both geometrical and in momentum, can be used to further separate electrons from jets after the application of isolation criteria. Apart from the momentum mismatch, jets or isolated pions distinguish from electrons by the shower shape which reflects in position mismatch and in other identification variables that can be built from track and cluster observables. Figure~\ref{fig:idvariables} (left) presents the E/p distributions obtained from the simulation of single electrons and QCD di-jets events.
As expected, jet induced showers don't match in energy with the track momentum.

In CMS, electron classes based on the different observed track-cluster patterns are used to optimize the electron identification using probability distributions  per class. As an illustration, the difference in $\phi$ between the cluster position and the track extrapolation from the outermost track hit is presented in Fig.~\ref{fig:idvariables} (right). The distributions observed for {\it golden} electrons and jets are clearly different. Not surprisingly, {\it showering} electrons which contains electrons with identified bremsstrahlung clusters are found more difficult to separate from jets. Further details can be found in \cite{CMSNOTEELE}. The electron identification is finally optimized by combining the different identification variables in a neural network or  in a likelihood function.

\begin{figure}
  \includegraphics[width=0.475\textwidth,height=.28\textheight]{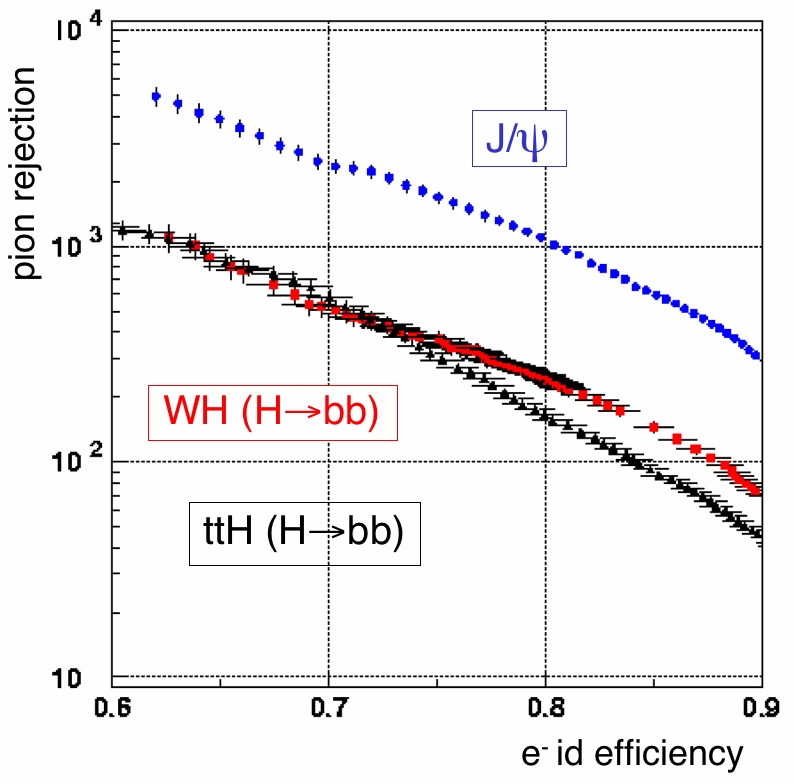}
  \label{fig:btag}
  \caption{Performances on single tracks of the ATLAS low $p_T$ electron algorithm. Pion rejection is shown as a function of the electron identification efficiency in J/$\Psi$ events, WH events and ${\rm t} \bar{{\rm t}}$H events with $H \rightarrow {\rm b} \bar{{\rm b}}$.}
\end{figure}

Finally, both ATLAS and CMS are considering the use of a dedicated algorithm for the identification and the reconstruction of electrons in jets. The ATLAS low $p_T$ electron algorithm starts from the track reconstruction and then looks for an energy deposition in the electromagnetic calorimeter around the extrapolated track position. The cluster properties are calculated and compared to  probability distribution functions or injected into a neural network. A similar procedure is being studied in CMS for the reconstruction of electrons in jets. Figure~\ref{fig:btag} shows the pion rejection obtained in ATLAS as a function of the electron identification efficiency for J/$\Psi$, WH and ${\rm t} \bar{{\rm t}}$ H events where $H \rightarrow {\rm b} \bar{{\rm b}}$. A 60$\%$ soft b-tagging efficiency is obtained for a rejection factor of 150 on WH events, which can be used to complement secondary vertex finding algorithms.

\section{Acknowledgments}
The author would like to thank D. Zerwas from the ATLAS Electron/Photon working group and C. Seez and Y. Sirois from the CMS ECAL/Egamma  working group for the fruitful discussions and  exchanges during this work. 

\bibliographystyle{aipproc}   % if natbib is available

\end{document}